# Magnetism of Sigma-Phase $Fe_{68}V_{32}$ Compound

[1]Piotr Konieczny[*] and [2]Stanisław M. Dubiel

[1]Institute of Nuclear Physics, Polish Academy of Sciences

ul. Radzikowskiego 152, 31-342 Kraków, [2]AGH University of Science and Technology, Faculty of Physics and Applied Computer Science, al. A. Mickiewicza 30, 30-059 Kraków, Poland

**Abstract**

**T**he magnetic phase diagram in the *H-T* coordinates has been determined for σ-$Fe_{68}V_{32}$ from the ZFC/FC magnetization measurements. The re-entrant character of magnetism, going from paramagnetic through ferromagnetic to spin-glass (SG) states, has been evidenced. The SG phase is magnetically heterogeneous, because two sub phases can be identified i.e. with the strong (SG-SI) and the weak (SG-WI) irreversibility. The $T_{irr}$ and $T_{cros}$ were quantitatively analysed using the mean-field theory and $\varphi_{irr}=1.6(2)$ and $\varphi_{cros}=0.91(9)$ values were obtained. A qualitative agreement with the Gabay-Toulouse model was reached. The isothermal magnetization measurements point to a soft magnetic behaviour of the studied sample. The *γ* critical exponent was determined with the Kouvel-Fisher approach yielding the value of *γ*=1.0(1) in line with the mean-field theory.

* Corresponding author: piotr.konieczny@ifj.edu.pl (Piotr Konieczny)

# 1. Introduction

The sigma-phase ($\sigma$) can occur in alloy systems with, at least, one transition element. Its crystallographic structure (space group $D^{14}_{4h}$-$P4_2/mnm$) is constituted by a tetragonal unit cell which accommodates 30 atoms distributed over 5 non-equivalent lattice sites. Characteristic features of $\sigma$ are high values (12-16) of coordination numbers and lack of stoichiometry [1]. Physical properties of $\sigma$ depend on the constituent elements, and, for a given choice of elements, on their mutual concentration. Concerning magnetic properties, the subject of the present study, they originally were revealed for $\sigma$ in the Fe-V [2] and Fe-Cr [3] systems. The magnetic ordering was for a long time regarded as a ferromagnetic one. However, recent studies revealed that it is much more complex as it has a re-entrant character, is weak and itinerant [4-7]. The Fe-V alloy system is exceptional regarding the existence of $\sigma$, as its field phase is the largest among the binary alloys. In particular, it can be formed in alloys whose V concentration ranges between ~31 and ~65 at. % and in a wide temperature range [8]. Thanks to the former, its magnetism, and in particular, the Curie temperature, $T_C$, could be tuned up to above the room temperature for a sample containing 33.5 at.% V [6] what is the record-high value for any $\sigma$. However, it follows from the crystallographic phase diagram of the Fe-V system [8] that the lowest concentration of V at which the $\sigma$-phase can be formed equals to ~31 at. %V. Thus the strongest magnetism of $\sigma$ should be observed in a $Fe_{69}V_{31}$ alloy. Thus the first aim of this study was to prepare such sample and determine its magnetic ordering temperature, $T_C$. Our second aim was to investigate its magnetic properties, and, particularly, to determine its phase diagram in the $H$-$T$ coordinates. For this purpose magnetization measurements were performed in a wide temperature and applied magnetic field ranges.

# 2. Experimental

## 2.1. Sample preparation

806.2 mg of the 99.999 pure vanadium and 1877.6 mg of the 99.99 pure iron were melted together under protective atmosphere of argon in an arc furnace. The melting process was repeated three times to obtain a more homogenous alloy. The mass of the alloy was unchanged by the melting process, hence the nominal concentration of vanadium i.e. x=32 at. % was assumed and used in the evaluation of measurements and discussion of results. The conversion of the alloy into the σ-phase was performed by an isothermal annealing at T=973K for 25 days. The validation of the α-to-σ phase transformation was done by recording a room temperature XRD pattern as well as a Mössbauer spectrum.

**2.2. Magnetic measurements**

Magnetic measurements were carried out with the use of MPMS XL SQUID (Quantum Design) magnetometer. A small sample (35.73 mg) of σ-$Fe_{68}V_{32}$ was mounted in the sample holder in such a way that a demagnetization effect was reduced. The diamagnetic contribution was estimated from magnetic measurements. Temperature dependences of magnetization in zero field cooled (ZFC) and field cooled (FC) conditions were recorded upon warming from 10 K to 380 K in an applied magnetic field between 20 Oe and 1000 Oe.

**3. Results and discussion**

The temperature dependence of ZFC/FC magnetization of the investigated sample (Fig. 1) can be used to determine three characteristic temperatures: (1) the magnetic ordering temperature, $T_c$; (2) the irreversibility temperature, $T_{irr}$; and (3) the crossover temperature, $T_{cros}$. Positions of these temperatures are marked with arrows in Fig. 2, which shows the ZFC/FC curves recorded in 500 Oe. The $T_c$ values were obtained independently with two methods viz. by the minimum of the first derivative of ZFC curve ($T_{c-dif}$) and with the modified Kouvel-Fisher approach ($T_{c-KF}$) [9]. In both cases the $T_c$ values are field independent with mean values of $T_{c-KF}$=335(2) K and $T_{c-diff}$=334(1) K, respectively. Noteworthy, the obtained value of the Curie temperature is in line with the one recently determined based on Mössbauer-effect measurements viz. 331 (1) K [10]. Moreover, the Kouvel-Fisher method was also used to determine the γ critical exponent. The obtained value of γ=1.0(1) points to the mean-field model and is smaller than previously found values for σ-$Fe_{52}V_{48}$ [4] or σ-$Fe_{66.5}V_{33.5}$ and σ-$Fe_{65.9}V_{34.1}$ [6].

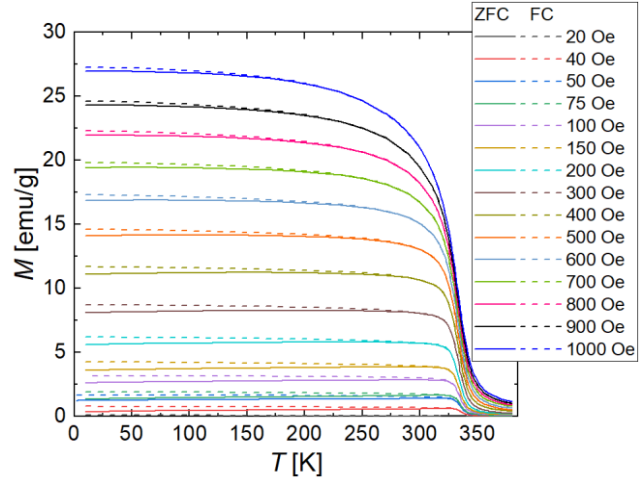

Fig. 1. Temperature dependences of magnetization measured in field cooled (dashed lines) and zero field cooled (solid lines) conditions.

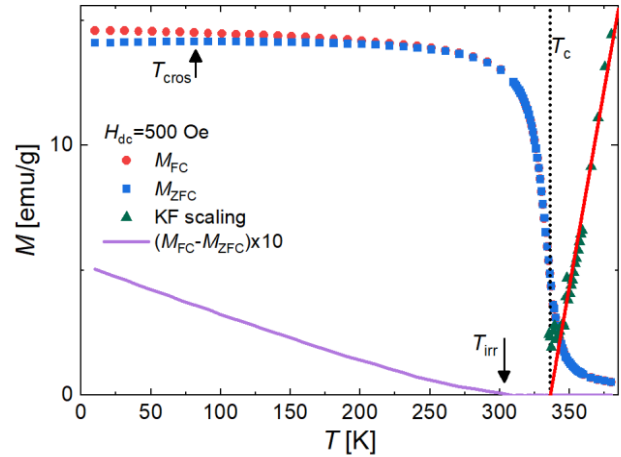

Fig. 2. Temperature dependence of ZFC (blue rectangles) and FC (red circles) magnetization for 500 Oe. The violet solid line shows the difference between FC and ZFC. The green triangles show the modified Kouvel-Fisher approach to the paramagnetic range data, while the red solid line stands for the corresponding best fit.

At $T_{irr}$, which is defined as the temperature of a bifurcation in the ZFC/FC curves, a spin-freezing process begins, while below this temperature irreversibility phenomena occur. The degree of irreversibility can be analysed qualitatively with the plots of a difference between FC and ZFC magnetization curves (Fig. 3) and quantitatively as the integrated area under $M_{FC}$-$M_{ZFC}$:

$$\Delta M = \int_{T_1}^{T_2}(M_{FC} - M_{ZFC})dT \qquad (1),$$

Where $T_1$=10 K and $T_2=T_{irr}$. It is evident from Fig. 3 and Fig. 4 that the irreversibility of the spin-glass (SG) state depends both on temperature and magnetic field. In all cases the degree of irreversibility increases with decreasing the temperature what can be understood in terms of a spin-freezing process. The field dependence is different. Namely, as shown in Fig. 4, $\Delta M$ increases steeply with $H$ for the fields smaller than ~100 Oe has a maximum around ~100-150 Oe followed by a graduate decrease.

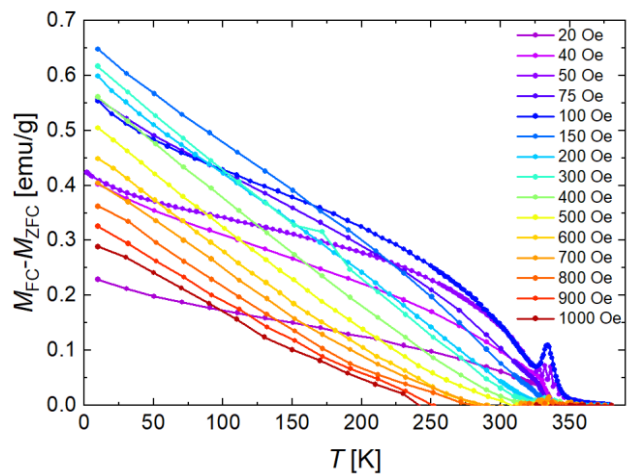

Fig. 3. Difference between FC and ZFC magnetization as a function of temperature. Solid lines are guides for the eyes.

Noteworthy, a similar behavior was observed for the σ-phase $Fe_{100-x}V_x$ alloys with x=33.5 and 34.1 [6], yet the maximum in the present case occurs at a higher field. This reflects, on one hand, a similarity in the magnetic properties of the three samples, but on the other hand a difference obviously due to a various concentration of vanadium.

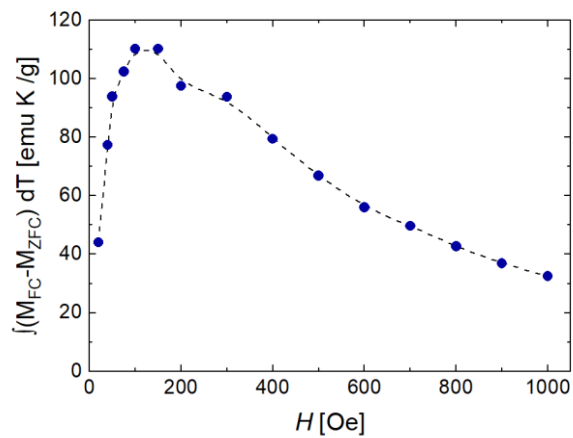

Fig. 4. Field dependence of temperature integrated difference between FC and ZFC magnetization. Dashed lines are guides for the eyes.

The $T_{cros}$ values were obtained as the temperature at which the $M_{ZFC}$ curves reveal a maximum and $T_{cros}$ is interpreted as a threshold between a weak and a strong irreversibility state of SG. All three characteristic temperatures obtained for different values of the applied field have been plotted in the $H$-$T$ plane to provide a magnetic phase diagram (Fig. 5). It is clear that the σ-$Fe_{68}V_{32}$ shows a re-entrant character of magnetism. Going from the high to low temperature, a transition from a paramagnetic (PM) over a ferromagnetic (FM) into a spin-glass state occurs. The ferromagnetic ordering was confirmed by the Curie-Weiss analysis which indicated a positive value of the Weiss constant Θ=334(1) K. Moreover, the ferromagnetic range enlarges as the applied field increases. The $T_c$ values are, as expected, $H$ independent. Instead, both $T_{irr}$ and $T_{cros}$ reveal an evident decrease with increasing the magnetic field. The $T_{irr}$ indicates the transition from FM to SG state, whereas the $T_{cros}$ divides the SG state into two sub states viz. with a weak (WI) and a strong (S) irreversibility. The characteristic temperatures of the SG state were quantitatively analysed using the mean-field theory, according to which the $T$-$H$ relationship can be described as:

$$T(H) = T(0) - aH^\varphi \qquad (2),$$

Where $T$ can stands for $T_{irr}$ or $T_{cros}$. Predicted values of $\varphi$, can be found throughout the literature with: (1) $\varphi$=2/3 known as the Almeida-Thouless line [11-13], (2) $\varphi$=2 known as the Gabay-Toulouse (GT) line [14-16] and $\varphi$=1 [17,18]. In the present case the best fit of $T_{irr}$ and $T_{cros}$ to Eq. 2 (Fig. 5) yielded $\varphi_{irr}$=1.6(2) and $\varphi_{cros}$=0.91(9), respectively. Only the $\varphi_{cros}$, within the error bars, agrees with the predicted value of $\varphi$=1, while the $\varphi_{irr}$ is lower than the $\varphi$=2 for the GT line. Nevertheless, a qualitative agreement with the Gabay-Toulouse model was reached, i.e. (1) existence of two characteristic lines within the spin-glass state, (2) both lines shift to lower temperatures with increasing the magnetic field and (3) the lines are characterized by different values of $\varphi$. Alternatively, the $T_{cross}$ – data can be successfully fitted with two straight lines (Fig. 5 bright green dashed lines) revealing different behaviour in low fields (up to 400 Oe) and high fields (above 400 Oe). Worthy of note, the latter approach (φ=1) perfectly agrees with the prediction by Dubiel et al. [17].

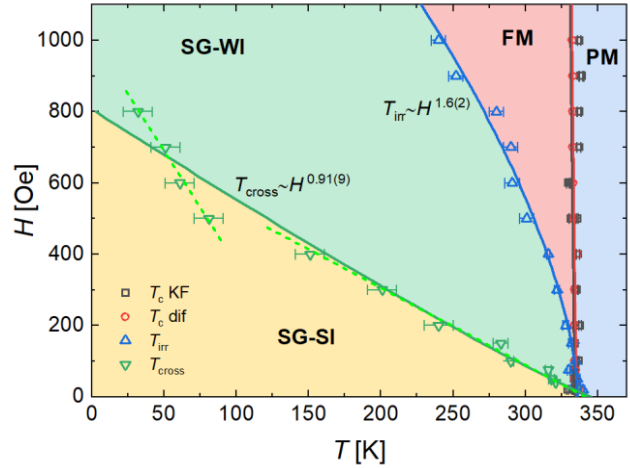

Fig. 5. Magnetic phase diagram for σ-Fe$_{68}$V$_{32}$ in the *H-T* plane. The solid and dashed lines represent the best fit to the corresponding data. The $T_{c\text{-KF}}$ and $T_{c\text{-dif}}$ were fitted with a linear function, while the $T_{irr}$ and $T_{cross}$ with equation (1). PM stands for the paramagnetic phase, FM for the ferromagnetic phase, SG-WI and SG-SI are the spin-glass states with the weak and the strong irreversibility, respectively.

The isothermal magnetization measurements (Fig. 6) in temperature range 2.0 – 300 K reveal a soft magnetic behaviour of σ-Fe$_{68}$V$_{32}$ with sharp increases in low fields (up to 1 kOe at 300 K and up to 3 kOe for lower temperatures) and small coercive fields at all temperatures (less than about 10 Oe). The magnetic measurements as a function of temperature and field can be used to determine the character of the magnetism in the studied compound with the Rhodes-Wohlfarth ratio [19], which compares the number of magnetic carries per atom derived from the Curie constant ($q_c$) and spontaneous magnetization ($q_s$) below $T_c$. In case of localized magnetic moments the ratio $q_c/q_s$ is of the order of 1, while for itinerant moments $q_c/q_s > 1$. The $q_c$ is determined in paramagnetic region from the total effective moment: $p_{eff}=\sqrt{q_c(q_c+2)}$ and the $q_s$ is the moment per magnetic atom of alloy at low temperature. Since in the studied case the paramagnetic range is narrow (between $T_c$ and 380 K, which was the maximum measured temperature), the Curie constant was only estimated from magnetization measurements as 229(17) K. The obtained $q_c/q_s \approx 5$ definitely indicates delocalized magnetism in the studied sample.

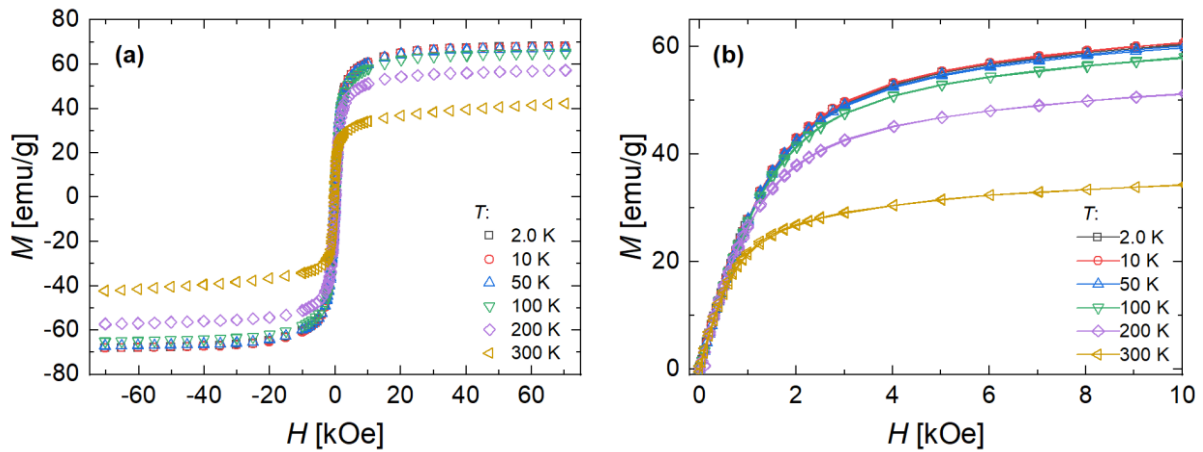

Fig. 6. Isothermal magnetization at 2, 10, 50, 100, 200 and 300 K in full (a) and reduced (b) magnetic field range. Solid lines are guides to the eyes.

## 4. Conclusions

The results obtained in this study permit the following conclusions to be drawn:

- The Curie temperature of the studied σ-$Fe_{68}V_{32}$ intermetallic compound has the value of 335(2) K which is the highest one ever reported for any σ-phase compound.

- The magnetism of σ-$Fe_{68}V_{32}$ has a re-entrant character i.e. on lowering temperature a transition from a paramagnetic over a ferromagnetic to a spin-glass state occurs.

- The spin-glass state is heterogeneous and it can be divided into two sub states viz. the one with a weak irreversibility and the other one with a strong irreversibility.

- The characteristic border lines predicted by the mean-field theory (MFT) have been identified and their behavior qualitatively agrees with the MFT predictions.

- The studied sample shows a soft magnetic behavior.

- The obtained Rhodes-Wohlfarth ratio testifies to an itinerant character of magnetism in σ-$Fe_{68}V_{32}$.

CRediT roles: **S. M. Dubiel**: Conceptualization; Sample preparation; Validation; Co-writing - original draft; **P. Konieczny:** Investigation; Formal analysis; Methodology; Validation; Visualization; Co-writing - original draft.

**Acknowledgements**

This work was financed by the Faculty of Physics and Applied Computer Science AGH UST and Institute of Nuclear Physics, Polish Academy of Sciences statutory tasks within subsidy of Ministry of Science and Higher Education, Warszawa.